\documentclass[twocolumn]{aastex631}

\usepackage{gensymb}
\usepackage{subfigure}

\begin{document}

\title{JWST/NIRCam Imaging of Young Stellar Objects III: Detailed Imaging of the Nebular Environment Around the HL Tau Disk}

\correspondingauthor{Camryn Mullin, Ruobing Dong}
\email{camrynmullin@uvic.ca, rbdong@uvic.ca}

\author[0009-0007-3210-4356]{Camryn Mullin}
\affiliation{Department of Physics and Astronomy, University of Victoria, Victoria, BC, V8P 5C2, Canada}

\author[0000-0001-9290-7846]{Ruobing Dong}
\affiliation{Department of Physics and Astronomy, University of Victoria, Victoria, BC, V8P 5C2, Canada}

\author[0000-0002-0834-6140]{Jarron Leisenring}
\affiliation{Department of Astronomy and Steward Observatory, University of Arizona, USA}

\author[0000-0001-7255-3251]{Gabriele Cugno}
\affiliation{Department of Astronomy, University of Michigan, Ann Arbor, MI 48109, USA}

\author[0000-0002-8963-8056]{Thomas Greene}
\affiliation{Space Science and Astrobiology Division, NASA’s Ames Research Center, M.S. 245-6, Moffett Field, 94035, CA, USA}

\author[0000-0002-6773-459X]{Doug Johnstone}
\affiliation{NRC Herzberg Astronomy and Astrophysics, 5071 West Saanich Rd, Victoria, BC V9E 2E7, Canada}
\affiliation{Department of Physics and Astronomy, University of Victoria, Victoria, BC, V8P 5C2, Canada}

\author[0000-0003-1227-3084]{Michael R. Meyer}
\affiliation{Department of Astronomy, University of Michigan, Ann Arbor, MI 48109, USA}

\author[0000-0002-4309-6343]{Kevin R. Wagner}
\affiliation{Department of Astronomy and Steward Observatory, University of Arizona, USA}

\author[0000-0002-9977-8255]{Schuyler G. Wolff}
\affiliation{Department of Astronomy and Steward Observatory, University of Arizona, USA}

\author{Martha Boyer}
\affiliation{Space Telescope Science Institute, 3700 San Martin Drive, Baltimore, MD, 21218, USA}

\author[0000-0001-9886-6934]{Scott Horner}
\affiliation{NASA Ames Research Center, MS 245-6, Moffett Field, CA 94035, USA}

\author{Klaus Hodapp}
\affiliation{University of Hawaii, Institute for Astronomy, 640 N. A'ohoku Place,
Hilo, HI 96720, USA}

\author{Don McCarthy}
\affiliation{Department of Astronomy and Steward Observatory, University of Arizona, USA}

\author{George Rieke}
\affiliation{Department of Astronomy and Steward Observatory, University of Arizona, USA}

\author[0000-0002-7893-6170]{Marcia Rieke}
\affiliation{Department of Astronomy and Steward Observatory, University of Arizona, USA}

\author[0000-0002-6395-4296]{Erick Young}
\affiliation{Universities Space Research Association, 425 3rd St. SW, Suite 950, Washington, DC 20024, USA}


\begin{abstract}

As part of the James Webb Space Telescope (JWST) Guaranteed Time Observation (GTO) program “Direct Imaging of YSOs” (program ID 1179), we use JWST NIRCam's direct imaging mode in F187N, F200W, F405N, and F410M to perform high contrast observations of the circumstellar structures surrounding the protostar HL Tau. The data reveal the known stellar envelope, outflow cavity, and streamers, but do not detect any companion candidates.
We detect scattered light from an in-flowing spiral streamer previously detected in $\textrm{HCO}^+$ by ALMA, and part of the structure connected to the c-shaped outflow cavity. 
For detection limits in planet mass we use BEX evolutionary tracks when $M_\textrm{p}<2M_\textrm{J}$ and AMES-COND evolutionary tracks otherwise, assuming a planet age of 1 Myr (youngest available age). 
Inside the disk region, due to extended envelope emission, our point-source sensitivities are $\sim5$ mJy ($37~M_{\rm J}$) at 40 AU in F187N, and $\sim0.37$ mJy ($5.2~M_{\rm J}$) at 140 AU in F405N. Outside the disk region, the deepest limits we can reach are $\sim0.01$ mJy ($0.75~M_{\rm J}$) at a projected separation of $\sim525$ AU.

\end{abstract}

\keywords{Young stellar objects --- Exoplanet Formation --- Direct Imaging --- Infrared Imaging --- Star Formation -- Envelopes}


\section{Introduction} \label{sec:intro}
Forming planets -- called protoplanets -- are forged around young stellar objects (YSOs) in regions of dust and gas known as protoplanetary disks \citep{Williams2011}. 
In this paper, we focus on the protoplanetary disk surrounding the class I star HL Tauri (HL Tau), located in the Taurus star forming region 140~pc away. 

The HL Tau disk is still embedded in the stellar envelope, as is typical for a system of its age ($\sim$0.1~Myr, \citealt{Stephens2017}). The envelope environment has been shown to have several active features such as an outflow/cavity and in-flowing streamers which could be associated with accreting material from the envelope to the disk \citep[e.g.,][]{Garufi2022}. 

Long-baseline interfermotric observations with ALMA revealed the disk around HL Tau to have multiple rings and gaps at solar system scales \citep{ALMA2015} It is theorized that these gaps were formed by interactions between the disk and one or more young planets via gravitational perturbations \citep[e.g.,][]{Paardekooper2022}. A planet orbiting a star within a disk may clear a gap in its orbital path which acts as a barrier for inward drifting dust, resulting in a dust ring forming outside the gap \citep{Pinilla2012, Zhu2012}. \cite{Dong2015}, \cite{Dipierro2015} and \cite{Jin2016} have suggested that each of the three major gaps (12~AU, 30~AU, and 65-75~AU) could be opened by a $\sim$Saturn mass planet. \cite{Dong2018} proposed that a sub-Saturn mass planet at $\sim71$~AU could produce all three gaps if the disk viscosity is sufficiently low.

Making observations of young planets in disks is crucial to testing planetary formation theories. However, protoplanets are orders of magnitude fainter than their host stars and therefore difficult to detect. 
Now, the James Webb Space Telescope (JWST) provides an unprecedented opportunity to push the limits of detection in the infrared (IR) where protoplanets are expected to have relatively high contrast compared to their host stars. With its increased sensitivity compared to ground based instruments, JWST is expected to increase the number of detected protoplanets and further our understanding of planetary formation \citep{Green2005, girard2022jwstnircam, Rieke2023}.

As part of the JWST Guaranteed Time Observation (GTO) program “Direct Imaging of YSOs” (program ID 1179), we preform high contrast imaging on some of the first planetary formation environments to be observed by NIRCam. HL Tau was chosen to test NIRCam's capabilities for imaging YSO environments, due to its long history of past observations providing compelling evidence for ongoing planet formation processes.
NIRCam's superior sensitivity and imaging quality in the infrared produce images of the disk's surrounding envelope in unprecedented detail, allowing us to probe these regions and look for possible companions interacting with the nebular environment. 
In addition, we search for planets both inside and outside the disk and set detection constraints. If planets do exist in the HL Tau disk and are detectable, they would likely be the youngest planets yet observed directly.

\begin{table}[t!]
\caption{Key Parameters of HL Tau and its Disk}
\def\arraystretch{1.25}
\begin{tabular}{lcc}\hline
Parameter & Value & Reference  \\ \hline
RA (J2000)  & 04:31:38.425 & 1 \\
DEC (J2000)  & +18:13:57.242  & 1    \\
Distance [pc]  & $140$  & 2  \\
Age [Myr]  & $0.1$ & 4  \\
$M_\star$ [$M_\odot$]  & $1.3$ & 1 \\
Spectral Type  & K$5\pm1$ & 3 \\
disk $i$ [$^\circ$] & $46.72\pm0.05$ & 1 \\
disk PA [$^\circ$] & $138.02\pm0.07$ & 1 \\
\hline
\end{tabular}\\\vspace{0.2cm}
\tablenotetext{}{References: (1)  \cite{ALMA2015}. (2)\cite{Rebull2004}. (3) \cite{White2004}. (4) \cite{Stephens2017}. }
\label{tab:HLTau}
\end{table}

This paper is organized as follows. Relevant past observations of HL Tau are summarized in section~\ref{sec:past observations}. In section \ref{sec:Obs and Data Red}, we break down how we took observations and our data reduction methods. In section \ref{sec:results} we show results and in section \ref{sec:discuss} we discuss them. Our summary and conclusions are in section \ref{sec:conc}.

\subsection{Summary of Past Observations} \label{sec:past observations}
Numerous studies of HL Tau have observed its surrounding environment using a variety of instruments ranging from the optical, through the infrared, and into sub-mm/mm wavelengths. Here, we present a summary of relevant past observations  to lay the groundwork for the new research we hope to accomplish. Key parameters of HL Tau and its disk are listed in Table~\ref{tab:HLTau}.

\subsubsection{Envelope Environment} \label{sec:Hubble}
Early images of HL Tau -- taken with the 2.2 m telescope at the Calar Alto Observatory, and the Infrared Telescope Facility (IRTF) -- revealed that the star was surrounded by a $\sim20\arcsec$ cloud of gas \citep{ Mundt1983, Grasdalen1984}. 
\cite{Mundt1983} discovered an ionized jet originating from the HL Tau region extending northeast at a position angle of $36\degree$, blueshifted from the star. 
\cite{cohen1983} observed an excess in the IR, which \cite{Grasdalen1984} hypothesized could result from re-radiation of 
starlight
absorbed by dust grains in a disk. These observations also suggested the star was surrounded by a spherical envelope. 

\cite{stapelfeldt1995} used the Hubble Space Telescope (HST) to image HL Tau at optical wavelengths. The images revealed a jet and structures known as Herbig-Haro objects, providing a glimpse of the active surrounding environment.  
\cite{Close1997} imaged HL Tau using the Canada France Hawaii Telescope (CFHT), and detected similar envelope features in H$'$, J, and K bands, as well as evidence for an active accretion disk with bipolar cavities. \cite{Close1997} surmised the upper and lower cavities were opened by an outflow.
\cite{Murakawa2008} imaged HL Tau using the AO-equipped near-infrared camera CIAO on Subaru. Their observations revealed a ``butterfly-shaped" polarization disk and extended envelope structure out to 4$\arcsec$ with a North facing extended feature.
\cite{Garufi2021, Garufi2022} studied the disk-outflow with ALMA, where SO and $\textrm{SO}_2$ molecules were found to spiral towards the star.  
\cite{Garufi2022} found a blueshifted infalling component in the NE direction, 
and a redshifted infalling component in the SW direction. The NW portion of the disk showed a redshifted component associated with a streamer.
From this, \cite{Garufi2022} determined that the presence of SO and $\textrm{SO}_2$ molecules can be used to probe accretion shocks in the disk, since such molecules correspond to intersections between the disk and in-flowing streamers.

\subsubsection{Disk Gaps and Planet Detection Limits} \label{sec:Disk}
The \cite{ALMA2015} Long Baseline Campaign provided the clearest images of HL Tau's disk structure to date, revealing multiple solar-system scale rings and gaps at mm wavelengths.
\cite{Yen2019} used ALMA to study $\textrm{HCO}^+$ emission from HL Tau, and found a gas gap at 30au consistent with being opened by a planet of $0.5-0.8~M_{\rm J}$, using the gap depth {\it vs} planet mass formula in \cite{Kanagawa2015}. 
In addition, \cite{Yen2019} detected a one-arm spiral in $\textrm{HCO}^+$emission, $\sim530$ AU in length extending from disk midplane and originating from an inflalling streamer.

The HL Tau disk has also been studied in the infrared prior to JWST. \cite{Testi2015} used the Large Binocular Telescope Interferometer (LBTI) LMIRCam \citep{Leisenring2012} to search for giant planets in HL Tau's outer 64 AU and 73 AU gaps.
\cite{Testi2015} took L$'$ ($\sim3.8~\mu$m) and K ($\sim2.2~\mu$m) band images, finding that the scattered light from the envelope impacted their ability to search for planets -- especially in K band.  With an inner masked region of 0\farcs18, \cite{Testi2015} did not detect any companion candidates in their images. 
The L$'$ NaCo-ISPY survey \citep{Cugno2023} also imaged HL Tau, but no companions were detected in the system.

\begin{table*}[t!]
\centering
\caption{Summary of Observations}
\def\arraystretch{1.25}
\begin{tabular}{lllllllllllll}\hline
Target & Prog. ID & Filter & $\lambda_\mathrm{mean}$ & W$_\mathrm{eff}$ & Readout & SUB & $N_\mathrm{gr}$ & $N_\mathrm{int}$ & $N_\mathrm{dither}$ & $N_\mathrm{roll}$ & $t_\mathrm{tot}$  & FWHM \\
& & & ($\mu$m)  & ($\mu$m) & & & & & & & (s) & ($^{\prime\prime}$) \\ \hline
HL Tau & 1179 & F187N  & 1.874 & 0.024 & RAPID & SUB160 & 10 & 120 & 4 & 2 & 2680 & $0\farcs064$      \\
HL Tau & 1179 & F200W  & 1.990 & 0.461 & RAPID & SUB160 & 10 & 120 & 4 & 2 & 2680 & $0\farcs066$      \\
HL Tau & 1179 & F405N  & 4.055 & 0.046 & RAPID & SUB160 & 10 & 120 & 4 & 2 & 2680 & $0\farcs136$   \\
HL Tau & 1179 & F410M  & 4.092 & 0.436 & RAPID & SUB160 & 10 & 120 & 4 & 2 & 2680 & $0\farcs137$     \\\hline
Injected Point Source \\\hline
P330-E    & 1538 & F187N  & 1.874 & 0.024 & RAPID & SUB160 & 7  & 2 & 4 & 1 & 15.6   & $0\farcs064$       \\
P330-E    & 1538 & F200W  & 1.990 & 0.461 & RAPID & SUB160 & 3  & 2 & 4 & 1 & 6.7    & $0\farcs066$      \\
P330-E    & 1538 & F405N  & 4.055 & 0.046 & RAPID & SUB160 & 10 & 2 & 4 & 1 & 22.3   & $0\farcs136$   \\
P330-E    & 1538 & F410M  & 4.092 & 0.436 & RAPID & SUB160 & 3  & 2 & 4 & 1 & 6.7    & $0\farcs137$     \\
\hline
\end{tabular}
\label{tab:observations}
\end{table*}

\begin{figure}[htpb]%
    \centering
    {\includegraphics[width=.5\textwidth]{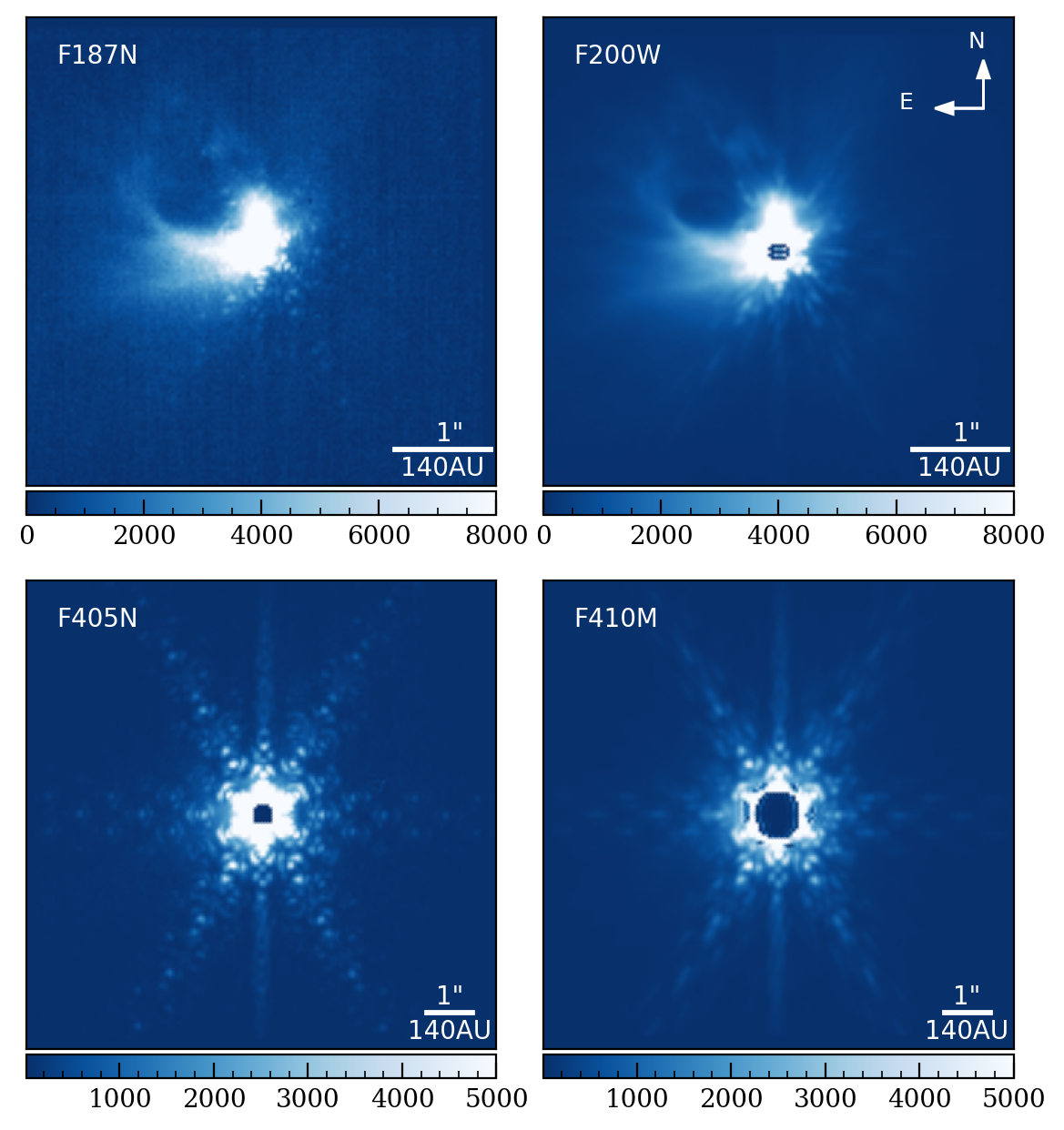}}
    \caption{Centered {\tt calints} files for the four filters used where color is in units of MJy/sr. Despite the PSF not being removed yet, the stellar envelope is apparent at short wavelengths and is as bright as the PSF. The central saturation regions extend out to 0\farcs1, 0\farcs2, and 0\farcs3 for the F200W, F405N and F410M filters, respectively, are masked out. The F187N observations do not have a saturated center. }
    \label{fig:centered_data}
\end{figure}

\begin{figure*}[htpb]%
    \centering
    {\includegraphics[width=0.95\textwidth]{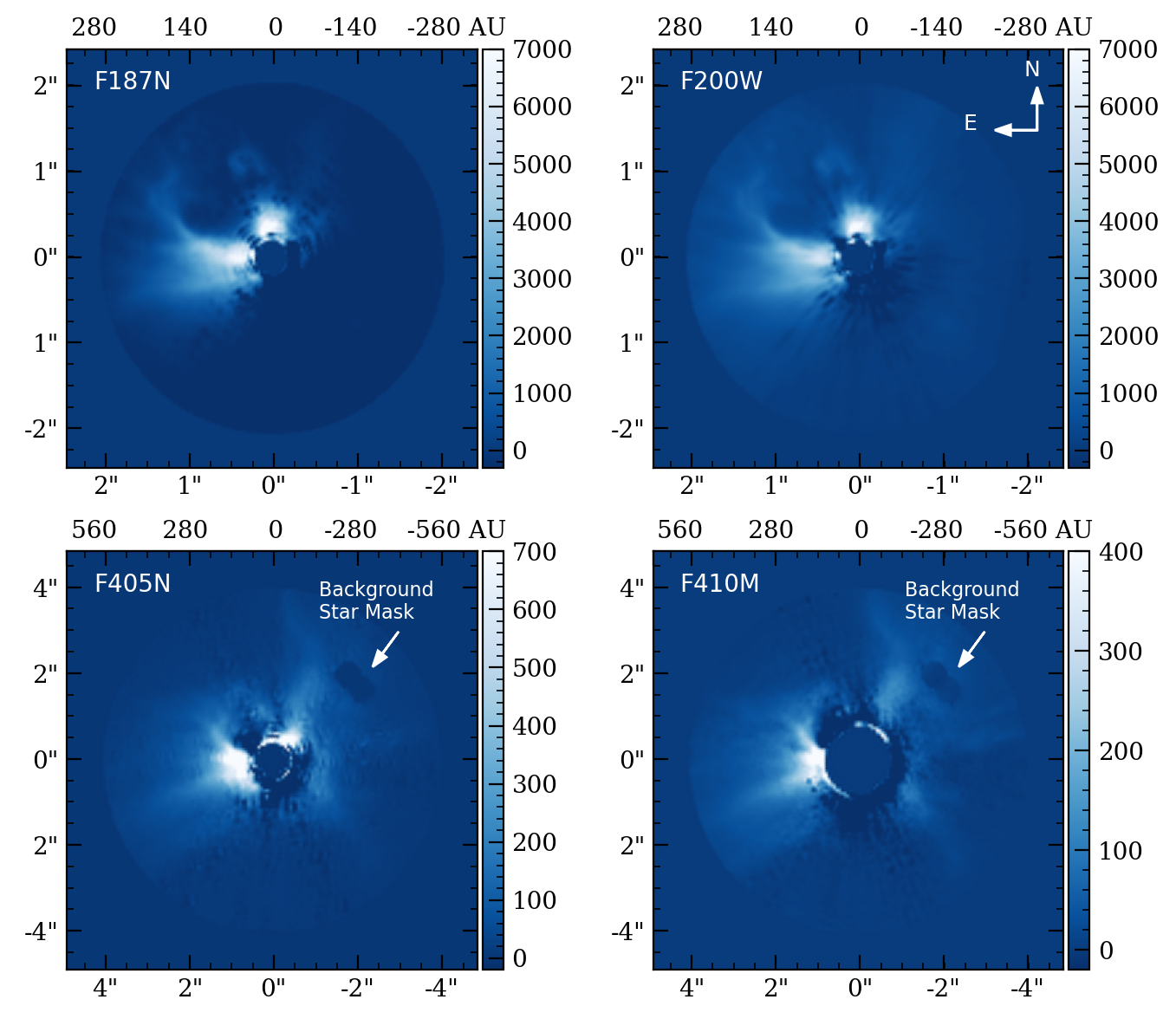}}
    \caption{PSF-subtracted data for all 4 filters where color is in units of MJy/sr. We utilize an annulus of inner radius 0\farcs2 for F187N and F200W, and 0\farcs4 for F405N. We use a larger 0\farcs8 mask for F410M -- which suffers from high levels of saturation -- to retain data in the outer envelope regions. The stellar envelope is the most prominent feature in our data. Negative spiral residuals out to a distance of $\sim1\arcsec$ have been introduced by using MWC 758 as a PSF reference. In addition, we apply a mask to an artifact introduced in the NW direction of the long wavelength images due to the presence of a background star in the MWC 758 data.}
    \label{fig:reductions}
\end{figure*}

\section{Observation and Data Reduction} \label{sec:Obs and Data Red}

Our observations of HL Tau were taken on UT 2022-09-29 with JWST/NIRCam in direct imaging (non-coronagraphic) mode  as part of GTO program 1179 (see Table~\ref{tab:observations}). This program will observe five disks in total, the other 4 being MWC 758 \citep{Wagner2024}, SAO 206462 \citep{Cugno2024}, PDS 70 (Leisenring et al, in prep), and TW Hya (observations pending).

For all observations, we used two filters for H line emission (F187N and F405N) and two for continuum emission (F200W and F410M). At each observing wavelength, we took images at two spacecraft roll angles -- $10\degree$ rotational separation -- to allow for angular differential imaging (ADI; \cite{Marois2006}). Four dither positions were used per image to correct for saturated or dead pixels. All integrations consisted of 10 groups in RAPID mode, resulting in 10 non-destructive reads up each integration ramp. We utilized the SUB160P subarray, setting 480 integrations 
per roll position for each filter, totaling 960 images and an exposure time of 2680s per filter. The centered images for each filter can be seen in Figure~\ref{fig:centered_data}.

We reduced the data using the standard {\tt jwst} pipeline (version 1.8.2 with crds version 11.16.15) 
for the initial reduction stages to obtain calibrated image files ({\tt calints} data), and then completed the reduction using a customized version of the open source {\tt PynPoint} data reduction pipeline \citep{amara2012, Stolker2019}.

\subsection{Official JWST/NIRCam Pipeline} \label{sec:JWST pipeline}
Following the same process as \cite{Cugno2024}, we began with the Level 1 {\tt uncal} files and ran stage 1 of the {\tt jwst} pipeline, which flags bad pixels, performs reference pixel correction (if available), corrects for non-linearity, and fits slopes to the ramp data to create {\tt rateints} files. We disabled the {\tt suppress\_one\_group} option to obtain signal information for pixels that saturated prior to sampling of the second group. Because of poor quality of the subarray dark calibration files, we turned off dark current correction. The jump detection threshold was set to 5 as per the suggestion of \cite{carter2023}. We then ran stage 2 of the pipeline which applies flux calibration and distortion corrections during the conversion of {\tt rateints} into {\tt calints}. 

\subsection{Customized PynPoint Pipeline} \label{sec:PYNPOINT}
Once the calibrated files were obtained, we completed all remaining reduction processes and post-processing steps in {\tt PynPoint} \citep{Stolker2019}. This pipeline managed bad pixel correction, centering, point-spread-function (PSF) generation, PSF subtraction, and synthetic planet injection. 

To correct for bad pixels, we followed the same process as \cite{Cugno2024} and replaced flagged pixels with the median value of the same sky location from the other three dither positions for the given filter and roll angle. It should be noted that the remainder of our reduction methods used for this source were slightly different than for the other objects from our program (MWC 758 and SAO 206462).

\subsubsection{Centering}
For centering purposes, we generated a perfectly centered model PSF using {\tt webbpsf} \citep{Perrin2014} by fitting a spectral energy distribution (SED) to HL Tau's photometry. 
HL Tau presents a particular challenge with centering, due to the asymmetric stellar envelope which is as bright as the PSF at the two shorter wavelengths, and cannot be replicated with {\tt webbpsf}. 
To mitigate this, we carried out a customized centering approach to handle these data's unique centering challenges. 

For a given filter, we first used cross-correlation to find the offsets of each integration relative to the position of the first integration within the first dither. The result provides the values necessary to shift all slope images to align with each other. In the first integration, we then masked the saturated inner core, and part of the bright envelope, accentuating the six diffraction spikes of the PSF as the dominant feature in the images. To improve accuracy in finding the PSF center, we applied a high-pass filter to the first image by convolving with a Gaussian kernel. The kernel size was dependent on the filter used and adjusted to optimize centering performance. 
We then masked the high-pass image in the same way as the original and masked corresponding regions in the simulated reference PSF. 
Finally, we performed cross-correlation between the observed masked image and the masked and perfectly centered reference PSF. These offset values were combined with the previously calculated relative offsets for each image to produce the final set of aligned and centered images (as seen in Figure~\ref{fig:centered_data}). 

\subsubsection{PSF Subtraction} \label{sec: psf sub}

To prepare for PSF subtraction, we masked the images, keeping an annulus of inner and outer radius 0\farcs2-2\farcs0 for the two short wavelength filters, and 0\farcs3-4\farcs0 for F405N. For F410M where the inner saturation was more severe, we constrained the inner radius to 0\farcs8 and kept the outer radius to 4\farcs0. Applying these masks removes the saturated inner core (along with pixels heavily afflicted by associated charge migration), and outer edges of the image that will appear after de-rotation. 

We subtracted the PSF using a technique known as Principal Component Analysis (PCA; \citealt{amara2012, Soummer2012}). The standard method used for the other targets in this GTO program was to use one telescope roll angle as the target image, and the other angle as the PCA PSF reference. This is a variation of classical ADI. While effective for some of the other targets, this is ineffective for HL Tau. The stellar envelope is bright at near-IR wavelengths, and azimuthally extended, which causes severe self-subtraction between the two roll angles.
To mitigate this, we used another program target, MWC 758, as the PSF reference.
This is similar to the technique Reference Differential Imaging (RDI), though no reference stars were imaged during these observations to allow for an ideal star match. MWC 758 was observed using the same filters, number of integrations, and subarray as with HL Tau. The MWC 758 files used for PSF reference were {\tt calints} files, centered using {\tt PynPoint}. 

With the use of MWC 758 as a PSF reference -- as opposed to roll-subtraction -- our results greatly improved. The PSF was subtracted from the data after the use of $\sim8$ principal components for each filter.
A disadvantage to using MWC 758 as a PSF reference, is the difference in SED shape between HL Tau and MWC 758. In addition to the two stars having different spectral types, HL Tau is 
partially obscured by its envelope, which modifies its SED.
As a result, and because MWC 758 possesses its own disk, the residuals suffered from over-subtraction, leading to loss of usable signals at the inner-most separation. This over-subtraction likely hinders our contrast performance (especially at $<1\arcsec$ separation) which is discussed further in section \ref{sensitivity}.

\section{Results} \label{sec:results}
The results of our PSF subtraction for all filters can be seen in Figure~\ref{fig:reductions}.
The disk itself is not visible in any filters due to strong saturation in the inner-most angular separation of the data and obscuring of the envelope. We do not detect any planet candidates in our residuals. The envelope (section \ref{sec: env}) is the primary feature in our residuals showing clear and detailed structure in all 4 filters. Our point-source detection limits in sensitivity are discussion in sections \ref{sensitivity} and \ref{sec: planet}. 

\subsection{Envelope Detection} \label{sec: env}
The surrounding envelope of HL Tau is prominently detected in all 4 filters.
The extended stellar nebula contains many features of interest, including a North facing streamer extending past $4\arcsec$ and a NE c-shaped structure near the inner disk extending to around 1\farcs5. This structure appears to be part of a previously identified outflow cavity which will be discussed further in section \ref{sec: HST}. We also tentatively detect part of a SW spiral streamer at $4~\mu$m -- previously detected in the form of $\textrm{HCO}^+$ emission -- which is discussed in section \ref{sec: spiral}. In addition to the nebular features we detect a seemingly detached ``hook-shaped" feature along the North side of the outflow cavity. This feature appears at $\sim1.8\arcsec$ separation from the central star, and is distinctly detached from the cavity material in all four filters. This feature appears to have been detected before at different wavelengths, though it has never been explicitly mentioned (see Section \ref{sec: HST}).

\subsection{Synthetic Planet Injection} \label{P330E}
Due to high saturation of the star in our data, we utilized the same method as with SAO 206462 \citep{Cugno2024} to determine our point-source detection limits, where the PSF of the standard G star P330-E was used for synthetic planet injections.  
P300-E was observed by NIRCam (PID 1538) on 2022-08-29, using all available detectors and filters with the SUB160 subarray. 
The observations of this star are reported in Table~\ref{tab:observations}.
This choice of synthetic planet injection allows us to compute accurate photometric calibration limits, but prevents us from obtaining contrast estimates with respect to the central star. 
To obtain calibrated flux limits, we can take a
scaling factor applied to P330-E's PSF and combine with the flux of the standard star (20.17 mJy, 21.23 mJy, 5.81 mJy and 5.84 mJy for F187N, F200W, F405N and F410M, respectively, Rieke et al., submitted to AJ).  
\begin{figure}
    \centering
    \subfigure{\includegraphics[width=.4\textwidth]{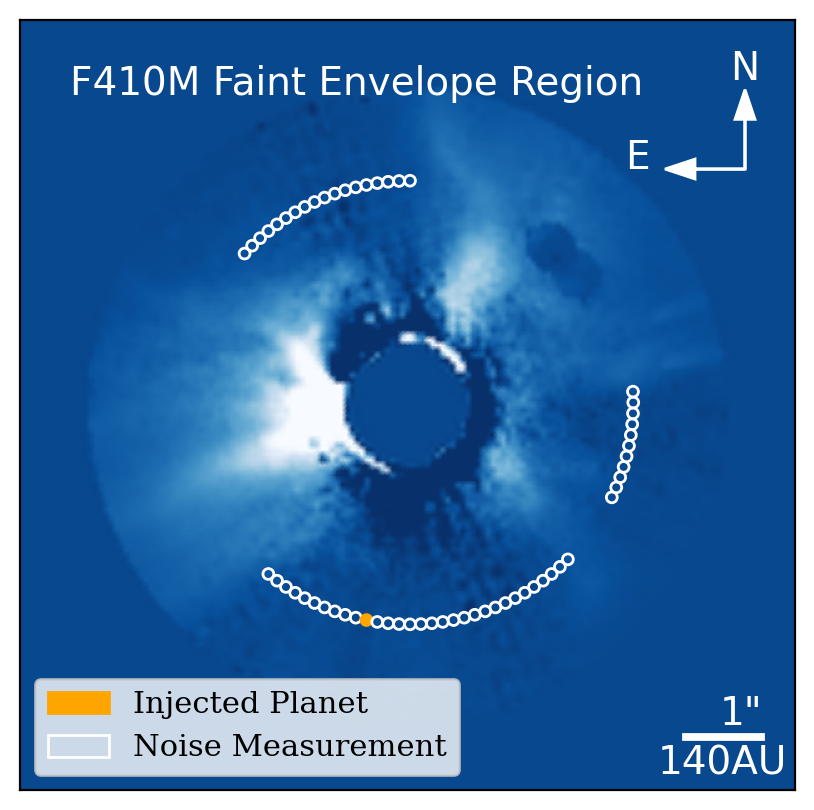}}
    \caption{Example of how we inject signal and noise 1~PSF FWHM diameter circular apertures depending on the position angle of the injected planet. A companion injected in an area of low envelope flux will only utilize apertures in low flux regions as a means of accurately measuring the noise.}
    \label{fig:Apperatures}
\end{figure}
\label{sec:companions}
\begin{figure*}[htpb]%
    \centering
    \subfigure[Sensitivity]{\includegraphics[width=.48\textwidth]{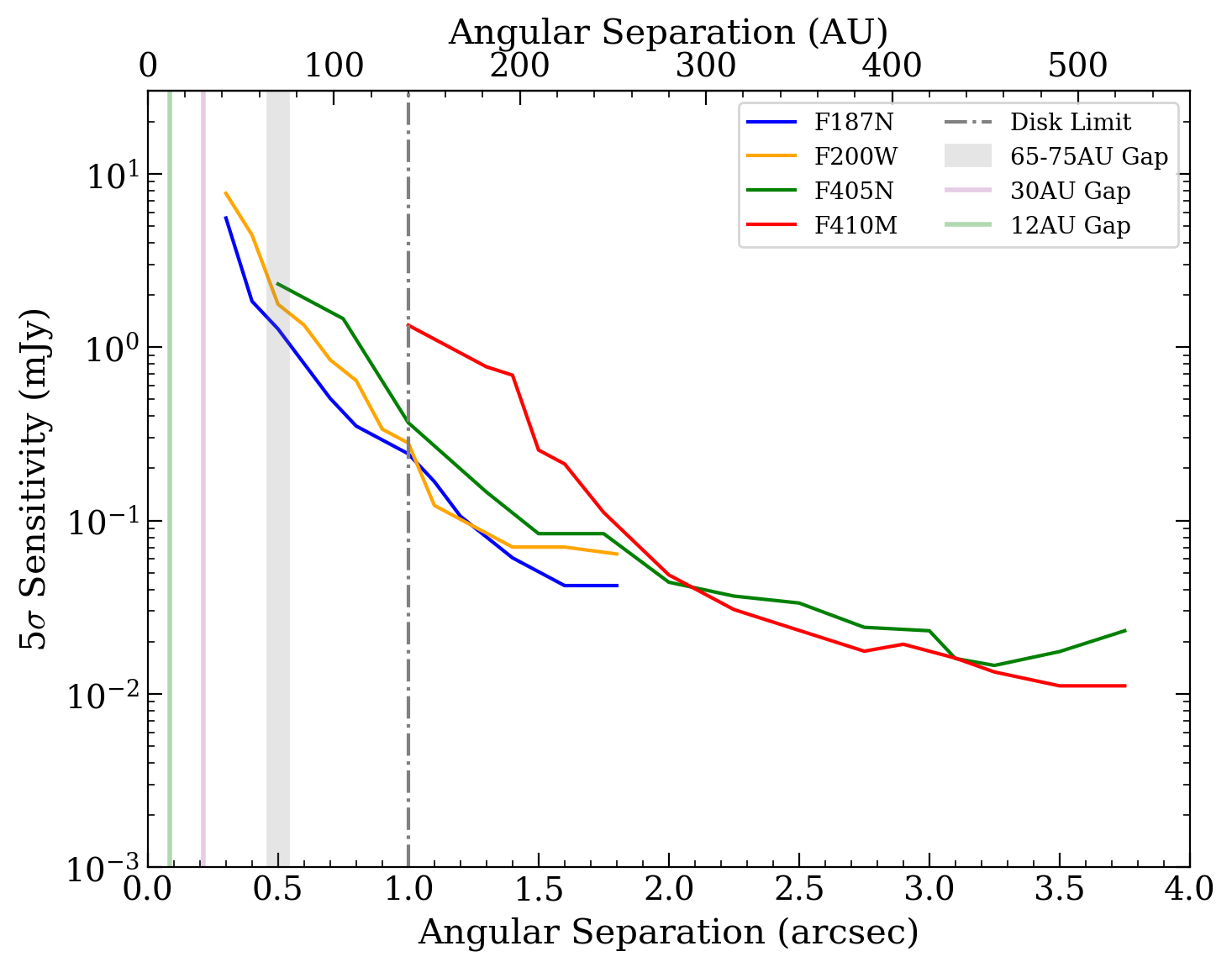}}
    \subfigure[Mass Limits]{\includegraphics[width=.47\textwidth]{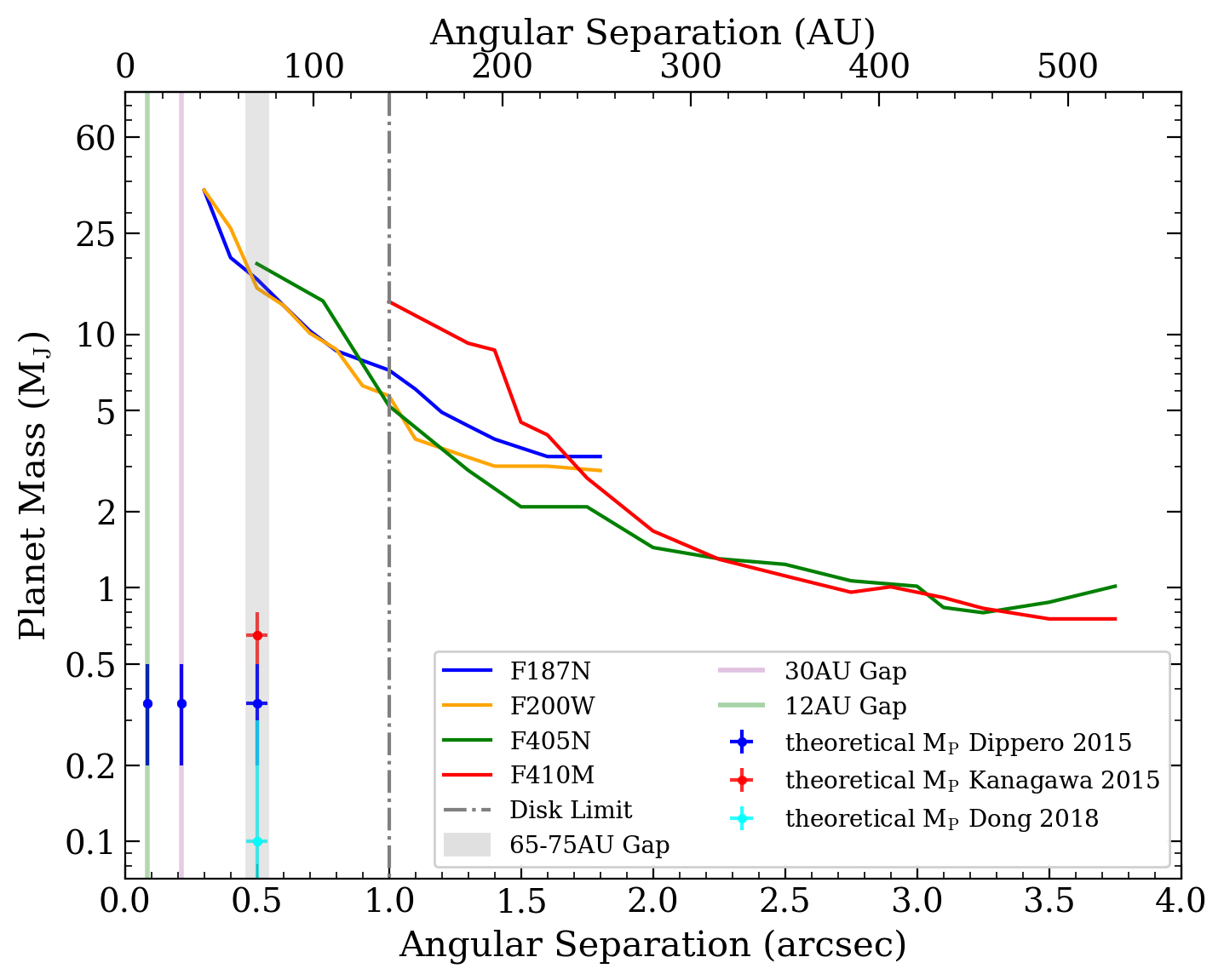}}
    \caption{(a) NIRCam $5\sigma$ sensitivity limits for HL Tau as a function of separation from the central star. 
    These values are only for regions with faint or no envelope detection. The edge of disk -- $\sim1\arcsec$ as imaged by \cite{ALMA2015} -- is represented by the vertical dash-dot grey line. The major disk gaps possibly hosting planets are plotted as vertical shaded regions. Due to saturation, only the 65-75AU gap is within our unmasked region. (b) Mass limits using BEX evolutionary track models when $M_{\rm p} < 2 M_{\rm J}$ and AMES COND models for larger masses. In the low envelope flux areas we reach limits as deep as $5.2M_{\rm J}$ at the disk edge in the F405N filter (assuming planets are 1Myr old). Our deepest limits are $\sim0.75M_{\rm J}$ out to $4\arcsec$. Points showing the approximate masses of the theoretically predicted planets are plotted in the major disk gaps to provide context.} 
    \label{fig:contrast}
\end{figure*}

Due to the presence of the envelope, sensitivity estimates for this source vary by angular separation from the star and position angle around the star. The regions dominated by envelope flux cannot be used to calculate $5\sigma$ sensitivity since the envelope dominates over background noise. Therefore, when calculating for $5\sigma$ sensitivity we only consider areas where the envelope readings flatten out and fluctuate around zero (comparable to background noise).
We determine high and low envelope flux areas by computing an azimuthal noise profile for each angular separation from the central star. To determine the baseline noise value in a low envelope flux region, we take an area in our reductions with low envelope flux and compute the standard deviation of the independent noise apertures in that area. We then define all apertures with a noise value greater than $2\sigma$ from the base-line noise to be in an area of high envelope flux and exclude those from our S/N measurements.
\begin{figure*}[htpb]%
    \centering
    \subfigure{\includegraphics[width=.85\textwidth]{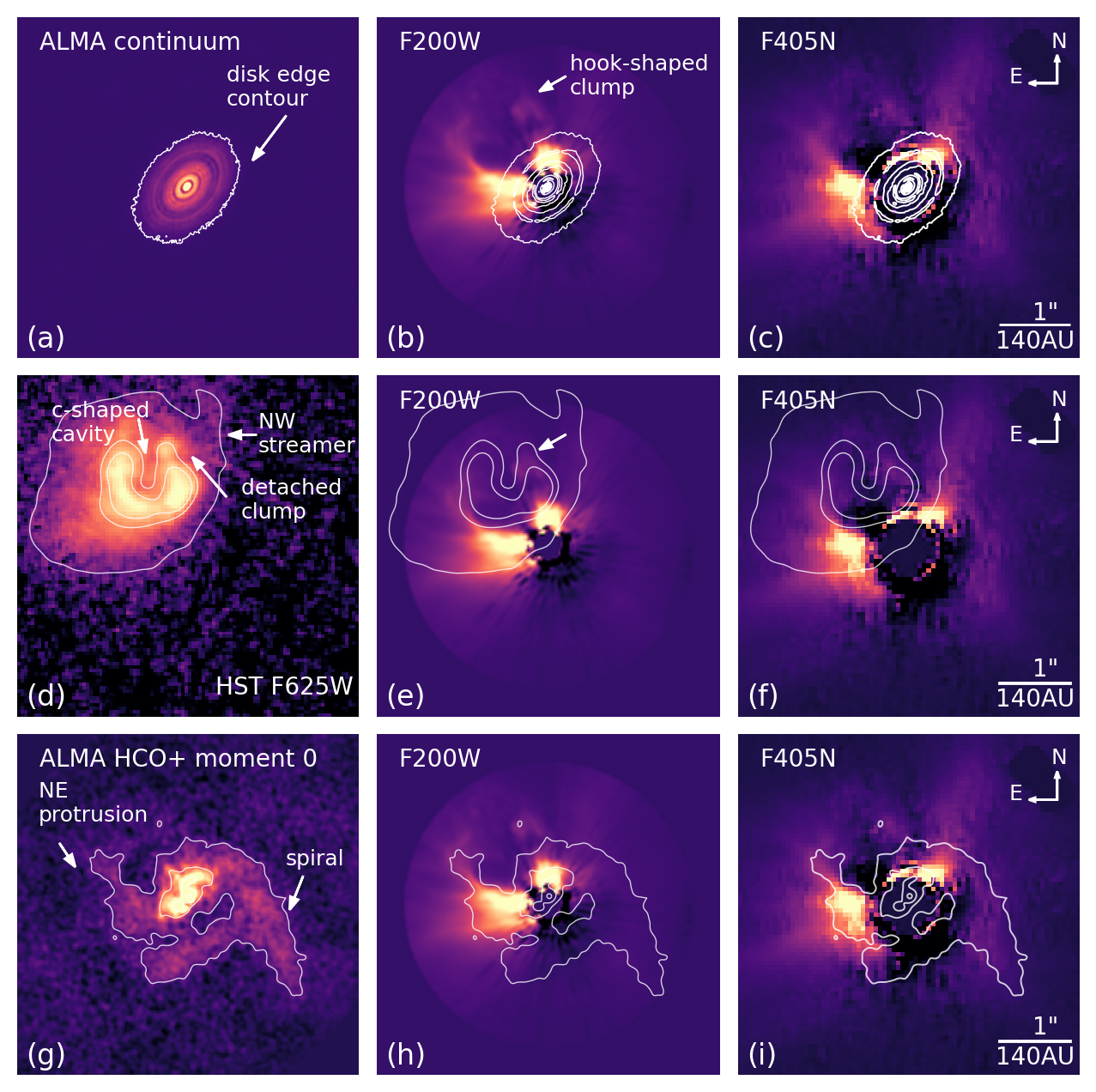}}
    \caption{Comparing features in our data with observations taken at different wavelengths. In all panels the star is located at the center, and the stellar position from the other datasets were manually aligned.
    \textbf{\emph{Top}}:
    (a) \cite{ALMA2015} mm continuum emission.
    (b) F200W residuals overlaid with ALMA contours to showcase the disk and gap locations. The disk is not visible to NIRCam. A notable hook-shaped clump is visible in our data a the edge of the outflow cavity.
    (c) F405N residuals overlaid with ALMA contours. 
    \textbf{\emph{Middle}}: 
    (d) 0.6$\mu$m HST data from PID9862 (log scaled). 
    (e) Comparison to F200W. The c-shaped outflow cavity is broader at 2$\mu$m. The white arrow points to the same hook-shaped clump seen in panel b. 
    (f) Comparison to F405N. The cavity is broader, and the upward stream is shifted west.
    \textbf{\emph{Bottom}}: 
    (g) ALMA $\textrm{HCO}^+$ moment 0 map from \cite{Yen2019}. 
    (h) Comparison to F200W. The spiral is not clearly detected at this wavelength, but the NE protrusion clearly aligns. 
    (i) Comparison to F405N. We detect a structure which appears to coincide with the HCO+ spiral.
    }
    \label{fig:Data Compare}
\end{figure*}
\subsection{Sensitivity Limits} \label{sensitivity}
We measure sensitivity by placing a circular aperture with diameter equal to the FWHM of the P330E PSF at the position of the P330E injection. We take the sum of pixel values in the aperture to give a flux measured in Jy. Then, at the same angular separation from the star, we place apertures in positions of negligible envelope flux
and take the standard deviation of these apertures to estimate the noise. An example showcasing the signal and noise apertures can be seen in Figure~\ref{fig:Apperatures}. Different scaling factors were applied to the injected P330E synthetic planet until $\sim5\sigma$ (defined here as $\textrm{S}/\textrm{N}=5$) was reached. 

Our sensitivity limits and corresponding mass limits in the low envelope flux regions can be seen in Figure~\ref{fig:contrast}. We note that due to both envelope obscuring and high saturation, NIRCam cannot detect the theorized Saturn mass planets at $\leq 70$ AU ($0.5\arcsec$) separation with $5\sigma$ confidence. 
At the edge of our $2~\mu$m FOV (1\farcs8 separation), our deepest limits are in the F187N filer, reaching $\sim40~\mu$Jy. At the edge of our $4~\mu$m FOV (3\farcs75) F410M reaches the deepest limits of $\sim10~\mu$Jy. Regarding the areas with high envelope signal, we inspected the data visually to look for planet signatures since we cannot quantitatively calculate $5\sigma$ sensitivity.

To estimate detection limits in planet mass, we used BEX \citep{Linder2019} evolutionary tracks for planets with $M_{\rm p} < 2 M_{\rm J}$ and AMES-COND models \citep{Chabrier2000} for more massive companions. Since no models were available for planets as young as 0.1 Myr, we chose the youngest available age of 1 Myr when interpolating companion masses. 
In these models, the older the planets, the more massive a companion with the same brightness must be to be detected. Because our system is younger than 1 Myr, and could have even younger planets, our mass estimates may be conservative since the limiting mass would decrease if the planets are younger. It should also be noted that these evolutionary models are less well constrained at young ages, and therefore the results for young systems are less robust.

In order to obtain mass limits in the disk regions of our data where planets are commonly searched for (0\farcs2 - 0\farcs8), we used a 0\farcs2 radius inner mask for the F187N, F200W and F405N filters. At these separations, we are sensitive to companions with $M_{\rm p}>8.6~M_{\rm J}$. 
At angular separations out to $1\arcsec$ -- the outermost edges of the disk -- we are sensitive to $M_{\rm p}\gtrsim7~M_{\rm J}$ for F187N, $M_{\rm p}\gtrsim5.7~M_{\rm J}$ for F200W, $M_{\rm p}\gtrsim5.2~M_{\rm J}$ for F405N, and $M_{\rm p}\gtrsim13~M_{\rm J}$ F410M, which suffers from the most over subtraction at inner angles.
We are most sensitive in regions outside the disk past $2\arcsec$ ($\sim$260 AU) where any remaining envelope obscuring is greatly reduced. At $2\arcsec$ in the low envelope flux regions we reach $\sim3~M_{\rm J}$  for the short wavelength filters and $\sim1.5~M_{\rm J}$ for long wavelength ones. Our deepest limits are $\sim0.75~M_{\rm J}$ at 3\farcs75 (525 AU) separation in the F410M filter.

\section{Discussion} \label{sec:discuss}

\subsection{Outflow Cavity} \label{sec: HST}
Past envelope imaging done by \cite{stapelfeldt1995} with HST and \cite{Close1997} with CFHT, revealed a c-shaped cavity produced by a jet extending NE $1\arcsec$ in length, centered 1\farcs2 from the central star. 
These observations have since been replicated with deeper imaging done by various telescopes. In turn, we resolve similar structures with NIRCam, and compare to 2004 HST observations from GO program 9862 as shown in the middle row of Figure~\ref{fig:Data Compare}.
The c-shaped outflow cavity seen with HST appears narrower at $0.6~\mu$m than with NIRCam. This is most clearly evident in panel (e) of Figure~\ref{fig:Data Compare}. While the cavity opens at approximately the same rotation angle, the structure which curls tightly inwards at $0.6~\mu$m, opens more widely at  $2~\mu$m and $4~\mu$m and extends outwards past 4$\arcsec$. This broadening of the c-shaped cavity at longer wavelengths is consistent with the \cite{Close1997} observations, and likely due to the change in optical depth with increased wavelength. The faint structure in panel (d) extending vertically North aligns with a previously identified streamer seen clearly at $4~\mu$m and faintly at $2~\mu$m. As evident by the HST contours, this feature shifts West at longer wavelengths, which is also consistent with \cite{Close1997}. 
 
The hook-shaped clump mentioned in section \ref{sec: env} is potentially detected in the HST data, which also shows a detached feature on the northern side of the c-shaped cavity. If this is the same feature, then it is further separated in the HST data ($\sim1.33\arcsec$). Whether this is due to different wavelengths, or a sign of radial shift over time is uncertain. 
However, when comparing the 2004 HST data at $0.6~\mu$m shown here with the results from \cite{Close1997} at $0.9~\mu$m and $1.6~\mu$m the clump consistently ``moves" inwards with increasing wavelength. Therefore it appears the clump position is wavelength dependant.
Material we associate with the clump is also faintly visible in the \cite{Yen2019} data, and is possibly at the same location as with JWST/NIRCam, as shown by the overlapping contour.

\subsection{Comparison to Past IR Observations}
Previous imaging with Subaru in the near IR by \cite{Murakawa2008} revealed the extended structure of the envelope in J (1.25~$\mu $m), H (1.63~$\mu $m) and K (2.2~$\mu $m) bands. These observing wavelengths would be most comparable to our F187N and F200W JWST observations. The NIRCam data match the Subaru observations, but with greater resolution and higher SNR than was possible in 2008. Our  observations in F405N and F410M
also showcase a similar extended envelope structure to 
\cite{Murakawa2008} -- such as an extended feature to the north ranging past $4\arcsec$ -- at different wavelengths to the Subaru observations. 

The \cite{Testi2015} infrared observations recovered some envelope structure in K band -- mainly the outflow around a position angle of 90$\degree$ -- and minimal signals around $300\degree$. They detected the same signals in L band, but less prominently in the residuals overall. Our reductions reveal the same basic envelope structure in these areas with the addition of more fine structures and details. 

\subsection{Streamers}\label{sec: spiral}
In the bottom row of Figure~\ref{fig:Data Compare} we compare our observations to the Moment 0 map by \cite{Yen2019}. As mentioned in section \ref{sec:Disk}, \cite{Yen2019} detected a SW extending spiral arm of infalling envelope in $\textrm{HCO}^+$. The spiral has a length of $3\arcsec$ starting from the North, then bending around the star, and extending to the SW, $2\arcsec$ away from the center. The same spiral appears to be visible in our data, as well as a NE extending feature tracing the c-shaped outflow cavity. We detect a portion of the spiral arm wrapping around the disk in F405N, and extending out to $3\arcsec$ at 4 $\mu$m. 
The position of our spiral aligns with the $\textrm{HCO}^+$ spiral, though the shape is somewhat different at IR wavelengths. The feature we detect traces a different component of the spiral than HCO$^+$ gas -- likely light scattering off dust tracing the spiral shape. 

Infalling streamers are likely continuously fed from the surrounding envelope, allowing the structures to survive for extended periods of time. Over the disk lifetime, a large streamer such as this with a mass infall rate of $\gtrsim5~M_\textrm{J}$\,Myr$^{-1}$ could greatly increase the mass available for planet formation \citep{Gupta2024}. If planet formation is ongoing in the disk, then this streamer is providing an influx of new material to the system for the planet(s) to accrete. The spiral streamer is not clearly detected in either of the short wavelength filters. Its absence can be seen in panel (e) of Figure~\ref{fig:Data Compare}. We do detect very faint structure at $2~\mu$m which traces the shape of the spiral seen at $4~\mu$m, but the inner part is not clearly visible. The dust and gas tracing the spiral streamer appear to have a composition such that they only peak at wavelengths $>2~\mu$m. The NW streamer is consistent with having a more robust detection at $>2~\mu$m, though it is more clearly visible at $2~\mu$m than the SW spiral, suggesting it may be composed of different material than the SW spiral. 

\subsection{Constraints on Planet Detection} \label{sec: planet}
Our sensitivity limits are hindered by high saturation at inner angles. It should also be noted that our saturation levels were not due to instrument error, but 10-20$\%$ higher throughput than was initially anticipated. 
As expected, we were unable to reach the depth needed to detect the theoretically predicted planets within the disk region -- sub-$M_{\textrm{J}}$ at $\leq1\arcsec$ from the star.
Our limiting factor is mainly the presence of the envelope. Young-hot planets peak in the near-IR (shorter wavelengths) and become fainter at longer wavelengths, making near-IR instruments optimal for such detections. We can also probe regions closer to the star at shorter wavelengths. However, HL Tau's envelope becomes more opaque at shorter wavelengths, making it difficult to 
effectively use the optimal wavelength range for young planet detection. 
It should be noted that while extinction from the envelope would factor into our ability to detect companions, we assumed all companions injected were foreground objects (not subject to extinction). Given this, the estimates listed here are the most optimistic case in terms of extinction. 

When comparing to IR imaging in the L$'$ band done by \cite{Cugno2023} and \cite{Testi2015}, our limits are comparable in the 4 $\mu$m filters if we use the same evolutionary tracks. Both studies also used similar data reduction methods to us -- a combination of ADI and PCA. We took sensitivity measurements both in the same inner region and further out than \cite{Testi2015} and \cite{Cugno2023} to a separation of 525 AU. While companions at this distance were not our primary target, 
it is still possible that a planet could exist outside the disk region as seen in recent papers by \cite{Cugno2024} and \cite{Pearson2023} where planet candidates were detected at wide separations from their stars. 

The \cite{Cugno2023} NaCo-IPSY survey ($3.8~\mu$m) reached $\sim9~M_\textrm{J}$ at $1\arcsec$ and $\sim5~M_\textrm{J}$ at $2\arcsec$ using our evolutionary tracks. Comparing to our low-envelope region observations in F405N we reached $\sim5.2~M_\textrm{J}$ at $1\arcsec$ and $\sim1.4~M_\textrm{J}$ at $2\arcsec$. Therefore, our observations reached a few $M_\textrm{J}$ deeper than VLT/NaCo. 
The \cite{Testi2015} LBTI/LMIRCam observations achieved an absolute magnitude of $\sim11.2$ at $1\arcsec$ -- corresponding to a mass limits of $\sim3.8~M_{\textrm{J}}$ using the evolutionary tracks adopted in this work -- which is deeper than our observations. These likely result from \cite{Testi2015} suppressing the envelope signal, allowing for a lower background limit. To subtract the envelope, they assume that no point sources would be detectable in K band ($2.2~\mu$m), and use the K band results to subtract the envelope from their L$'$ data. If a similar method were applied to our data, then we would perhaps reach deeper limits than are achieved here. 
Future work focused primarily on deep detection in embedded disks could explore this further.

Comparing our limits to other PID1179 targets, MWC 758,  SAO 206462 and HL Tau 
all manage to achieve a depth of $\sim1-10~\mu{\rm Jy}$ at $4\arcsec$. 
The depth reached in each filter relative to the others differed for each object. For MWC 758, the F405N filter consistently reaches deeper than F410M. For our data and SAO 206462, F405N reaches deeper at separations $\lesssim1$-$1.5\arcsec$, while F410M generally surpasses F405N at larger separations. This is likely due to saturation at the inner angles where the brighter-fatter effect and charge migration hinder sensitivity in F410M. 
At $<2\arcsec$ separation MWC 758 and SAO 206462 reach the deepest limits in F187N, while for HL Tau F187N generally reaches deepest in sensitivity but not in mass limit. 

\begin{figure}%
    \centering
    \subfigure{\includegraphics[width=0.5\textwidth]{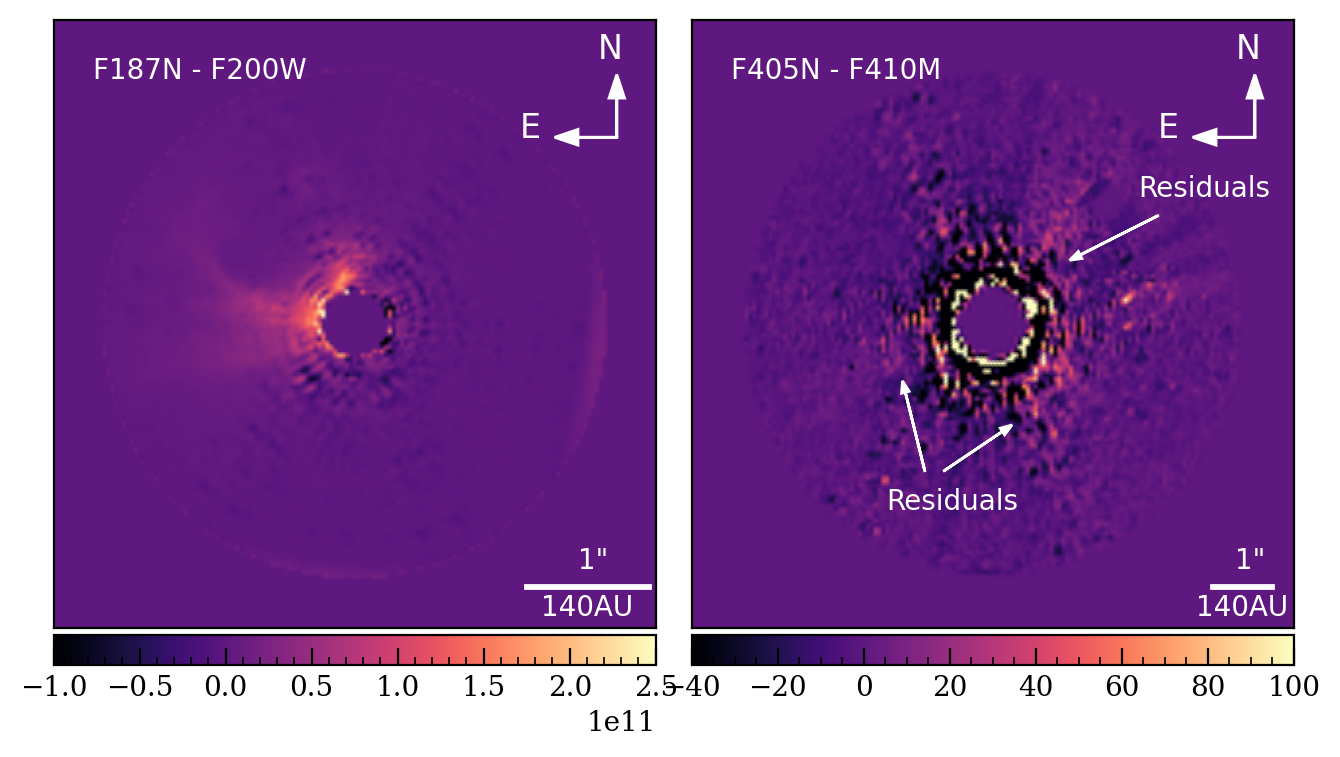}}
    \caption{Continuum subtracted residuals for F187N (Pa-$\alpha$) and F405N (Br-$\alpha$). The faint residuals shown in the left panel could indicate Pa-$\alpha$ accretion shocks in the disk region. There are also potential signs of Br-$\alpha$ accretion in the disk, though it is fainter when compared with Pa-$\alpha$.}
    \label{fig:accretion}
\end{figure}

\subsection{Accretion}
When comparing to past observations, we chose F405N to represent $4~\mu$m for Figure~\ref{fig:Data Compare} since the saturation of F410M obscures important features at $\leq1\arcsec$ separation. While the morphological differences we see in F405N -- as compared to ALMA and HST -- can be attributed to difference in wavelength, it should also be noted that F405N is centered around the Br-$\alpha$ line, and so other physics may be involved when comparing to that filter. To test for evidence of accretion we removed any continuum from F405N by subtracting F410M. As can be seen in the right panel of Figure~\ref{fig:accretion}, there are tentative signs of accretion close in the disk region, as evident by the
positive features in the residuals around the intersection between the disk and the NW and SW streamers. We performed a similar test with F187N, though we scaled each filter by the sum of the flux in the other (in the outer region less affected by PSF subtraction effects), since F200W is likely not an ideal representative of F187N continuum due to the difference in wavelength. It is possible the residuals seen in the left panel of Figure~\ref{fig:accretion} indicate Pa-$\alpha$ accretion shocks near the disk, though the signals may also be artificial due to the scaling factor. We see signals of the potential accretion at the location of the outflow cavity.
\cite{Garufi2022} previously detected signs of accretion shocks with SO and $\textrm{SO}_2$ in the NW disk region 
-- our residuals in the NW of F405N appear to align with this.

\subsection{Requirements to Detect Small Embedded Planets} 
Observing smaller planets within the disk region, which is highly obscured by the envelope, does not seem to be possible at NIRCam wavelengths. Observations performed at longer mid IR wavelengths may better bridge the gap between our NIRCam envelope observations and the ALMA disk observations. Longer mid-IR observations may better penetrate the envelope, while not losing emission from young-hot planets. However JWST/MIRI does not have the necessary angular resolution at $\leq 70~{\rm AU}$ to detect the theoretical $0.5-0.8~M_{\rm J}$ gap opening planets.
This type of observation is complicated with space-based instruments, since it requires both long wavelengths and high angular resolution. No current instruments have the resolution and wavelength range required for such an observation. However, probing planets in an embedded disk may be possible with future ground based observations using a new class of Extremely Large Telescopes (ELTs). The Mid-infrared ELT Imager and Spectrograph \citep[METIS,][]{Brandl2021} instrument on the upcoming European Extremely Large Telescope (ELT) has planet formation and circumstellar disks as one of its primary science goals. ELT will be a complementary telescope to JWST, since both operate at a similar wavelength range and have different strengths. While JWST can detect fainter objects at a larger separations, ELT/METIS can provide higher angular resolution and sharper images
 \citep{brandl2012, Brandl2021}. An ELT observation of HL Tau using METIS could nicely complement the JWST/NIRCam observation with the possibility of detecting gap-opening planets. 

Another upcoming extremely large telescope is the Thirty Meter Telescope (TMT). TMT will also operate in the mid-IR, and will have superior resolution to JWST. Simulations have shown that TMT can detect $\leq0.1~\rm{M}_{\rm J}$ planets at $\leq 10~{\rm AU}$ at a distance of 140 pc \citep{skidmore2015}. This resolution would be ideal for imaging the 12 AU, 30 AU, and 70 AU gaps in HL Tau. It should be noted that these limits (as with ELT) do not factor in the presence of an envelope, however at mid-IR wavelengths obscuring from the envelope would be less severe than as seen with NIRCam. As learned from our observations, future observations of this target or ones like it must factor in envelope flux in order to select the right instrument and observing modes when searching for planets in embedded disks.

\section{Summary and Conclusions} \label{sec:conc}
We made JWST NIRCam observations of HL Tau, a class I star with a multi ring and gap disk. We obtained images out to 4\arcsec (560 AU), with 4 NIRCam filters, and provided a more complete picture of the dust and gas surrounding this young star. Our findings are as follows: 
\begin{itemize}
    \item With NIRCam's sensitivity we were able to obtain the most detailed envelope structure for this system in the infrared. We detect HL Tau's protostellar envelope in all 4 filters as well as the c-shaped outflow cavity and NW streamer previously detected by \cite{stapelfeldt1995, Close1997, Murakawa2008}, and a detached hook-shaped clump (Figure~\ref{fig:reductions}, \ref{fig:Data Compare}).
    
    \item We detect part of an infalling streamer previously detected in the form of $\textrm{HCO}^+$ emission with ALMA \citep{Yen2019} in our 4 $\mu$m filters (Figure~\ref{fig:reductions}, \ref{fig:Data Compare}). The spiral streamer is most clearly detected in the F405N filter, which is centered on the Br-$\alpha$ line, however the spiral feature appears to be mainly scattered light from the continuum.

    \item We do not detect the ALMA dust emission protoplanetary disk since the envelope flux dominated over any disk light at 2-4 $\mu$m wavelengths. We also do not detect any protoplanet candidates.
    
    \item Our deepest detection limits within the ALMA disk region are $5.2~M_{\rm J}$ in the F405N filter -- deeper than VLT/NaCo. 
    At the edge of our FOV (4\arcsec) we reach $\sim0.75~M_{\rm J}$ (Figure~\ref{fig:contrast}).

    \item We see tentative evidence for Br-$\alpha$ accretion signatures where the disk intersects with streamers. We also see potential signs of Pa-$\alpha$ accretion at the intersection between the disk and the outflow cavity (Figure~\ref{fig:accretion}).
  
\end{itemize}

This work, along with \cite{Wagner2024}, and \cite{Cugno2024}, 
provide the first observations of young stellar objects with JWST/NIRCam. HL Tau is the youngest YSO imaged by this instrument, providing images of its young stellar environment in unprecedented detail.

\section{Acknowledgements} \label{sec:acknowledgements}
We thank the anonymous referee for the thoughtful and constructive questions and suggestions.
The authors are grateful for support from NASA through the JWST /NIRCam project, contract number NAS5-02105 (M. Rieke, University of Arizona, PI).  
C.M. and R.D. are supported by the Natural Sciences and Engineering Research Council of Canada (NSERC), the Alfred P. Sloan Foundation, and the Government of Canada’s New Frontiers in Research Fund (NFRF), [NFRFE-2022-00159].
K.W. acknowledges support from NASA through the NASA Hubble Fellowship grant HST-HF2-51472.001-A. 
D.J. is supported by NRC Canada and by an NSERC Discovery Grant.
We would like to thank Yen Hsi-Wei for providing us with moment 0 ALMA $\textrm{HCO}^+$ data from \cite{Yen2019}. This paper makes use of the following ALMA data: ADS/JAO.ALMA\#2011.0.01234.S. ALMA is a partnership of ESO (representing its member states), NSF (USA) and NINS (Japan), together with NRC (Canada), MOST and ASIAA (Taiwan), and KASI (Republic of Korea), in cooperation with the Republic of Chile. The Joint ALMA Observatory is operated by ESO, AUI/NRAO and NAOJ. We would also like to thank Karl Stapelfeldt for directing us to the 2004 HST data from PID9862. 
Some of the data presented in this article were obtained from the Mikulski Archive for Space Telescopes (MAST) at the Space Telescope Science Institute. The specific observations analyzed can be accessed via \dataset[DOI: 10.17909/erkx-v276]{https://doi.org/10.17909/erkx-v276}.

\bibliography{HLTau}{}
\bibliographystyle{aasjournal}



\end{document}